\documentclass[12pt]{iopart}
\usepackage[noadjust]{cite}
\usepackage{caption}
\usepackage{graphicx}
\usepackage{amsmath}
\usepackage{multicol}
\usepackage{hyperref}
\usepackage[normalem]{ulem}
\hypersetup{
     colorlinks   = true,
     linkcolor    = blue,
     citecolor    = red
}

\begin{document}

\title[Effects of Next-Nearest-Neighbor Hopping on the Low-Dimensional Hubbard Model]{
Effects of the Next-Nearest-Neighbor Hopping on the Low-Dimensional Hubbard Model: Ferromagnetism, Antiferromagnetism, and Superconductivity
}

\author{Luhang Yang $^{1,\ast}$, Adrian E. Feiguin $^{2}$, Thomas P. Devereaux $^{3,4}$, and Elbio Dagotto $^{1,5}$}

\address{1 Department of Physics and Astronomy, University of Tennessee, Knoxville, Tennessee 37996, USA}
\address{2 Department of Physics, Northeastern University, Boston, Massachusetts 02115, USA}
\address{3 Stanford Institute for Materials and Energy Sciences, SLAC National Accelerator Laboratory, 2575 Sand Hill Road, Menlo Park, CA 94025, USA}
\address{4 Department of Materials Science and Engineering, Stanford University, Stanford, CA 94305, USA}
\address{5 Materials Science and Technology Division, Oak Ridge National Laboratory, Oak Ridge, Tennessee 37831, USA}

\ead{luhangyang@utk.edu}

\vspace{10pt}

\begin{abstract}
    The Hubbard model has attracted considerable interest due to its prototypical role in describing strongly interacting electronic systems, such as high-critical-temperature superconductors as well as many novel quantum materials. By introducing next-nearest-neighbor (NNN) hoppings to the Hubbard model, the phase diagram becomes richer, and fascinating phenomena arise in both, one-dimensional chains and square lattices, such as:  antiferromagnetism (AFM), ferromagnetism (FM), superconductivity (SC), as well as charge orders, among others. 
    Moreover, NNN hoppings play a fundamental role in understanding effects of doping on magnetism and pairing orders in strongly interacting regimes. 
    In this article, we review the recent progress in understanding the different competing phases of this model in one and two dimensions from a computational perspective. We comment on the pressing technical challenges, illustrate the controversial results concerning the emergence of the SC phase, and conclude with our perspectives on future explorations.
\end{abstract}

\maketitle

\section{Introduction}

An adequate microscopic model capturing high-temperature superconductivity is crucial for the theoretical comprehension of cuprate  superconductors. In addition, it can also provide a template to guide experimental explorations into other quantum materials with similar characteristics. 
Much of the research in this direction has been guided by two questions: how to reduce the number of degrees of freedom to those that play a fundamental role in the physics of the cuprate materials, and how to represent the electron interactions. For instance, it is conventionally assumed that the fundamental physics occurs in the two-dimensional copper-oxygen planes, and that the apical oxygens do not play a relevant role. This has allowed theorists to postulate a three-band ``Emery'' model that only includes the $d_{x^2-y^2}$ orbital of the copper, and one $p$ orbital per oxygen \cite{Emery_1,VARMA1987681}. This model, in turn, can be reduced to a one-band effective model by means of the Zhang-Rice singlet construction \cite{ZRS}.
The single-band Hubbard model and its extended forms are the simplest models that incorporate the lattice potential and electron interactions, driving many experimental efforts with cold atomic gases \cite{Jordens2008,Esslinger2010,Endres2011,Georgescu2014,Hart2015,Duarte2015,Boll2016,Cheuk2016,Brown2017,Hilker2017,Mazurenko2017,Gross2017,Scherg2018,Salomon2019,Nichols2019,Chiu2019,Koepsell2019,Vijayan2020,Koepsell2021,Sompet2022,Hirthe2023,Hartke2023,Kohlert2023,Bourgund2025,Xu2025,Chalopin2025,Nielsen2025}. 
Under this framework, an important question arises: \textit{how does the non-interacting band structure interplay with the electron interactions}.

To address this question, we need to first reach a proper description for the non-interacting bands. Studies have revealed that next-nearest-neighbor (NNN) hopping beyond the adjacent sites is critical 
for describing the superconducting phase
\cite{absence_hubbard_2D,tasaki1998hubbard}. 
This yields a simple yet rich Hamiltonian (\ref{t1t2U}) that accommodates various quantum orders of great interest, including superconductivity (SC), antiferromagnetism (AFM), ferromagnetism (FM), and charge/stripe orders. On square lattices, the Hubbard model with NNN hopping has been widely adopted as the paradigmatic model to deliver a theoretical understanding of cuprate superconductors.
The electron-hole asymmetry, the magnetism in the under-doped regime, and the SC phase on the electron-doped side of the single-band Hubbard model (with NNN hopping), are all in good agreement with the experimental results of cuprate superconductors \cite{hubbard_2d_SC_in_both,antiphase_huang2017numerical,Tmatrix_hole_no_SC, raghu_review,positive_t2_enhance_AFM_ED,electron_afm_1,Elbio_tJ_2leg_ladder_pairing,Hubbard_2leg,jiang2019sc_4leg_negativet2,Hubbard_positive_t1_t2_4leg,plaqutte_SC_4leg,shengtao1_t1_t2_J_6_8_legs_SC_on_electron_doped_side,shengtao2_8_leg,white_enhanced_SC_electron_side_8leg,dona1_6_8_SC_both_electron_hole_doped,dona2,yifan_6leg_SC_electron_doped_side,yifan2_6leg_SC_electron_side}.
In this context, it is crucial to understand how the introduction of the NNN hopping changes the physics of the Hubbard model.

The effect of the NNN hopping in the non-interacting limit is clear: it will alter the Fermi surface geometry to the point in which one may realize hole pockets. In one spatial dimension the two Fermi points are replaced by four (Fig.\ref{fig:structure} (c)), and in two spatial dimensions the perfect Fermi surface nesting is eliminated. As a consequence, the NNN hopping makes the bands flatter at high or low energies ((Fig.\ref{fig:structure} (d)-(m))), depending on the sign of the hopping coefficient. 
By turning on interactions, however, the effect of the NNN hopping is more ambiguous, especially when the system has intermediate to strong interactions. 
One would expect that the narrow bands together with interactions can eventually give rise to superconductivity or FM phases in some parameter regimes. Nevertheless, the quantum fluctuations induced by the interplay between the electronic bands and interactions make the conclusions sensitive to many factors, and results that are in some cases contradictory \cite{jiang2019sc_4leg_negativet2,Hubbard_positive_t1_t2_4leg,plaqutte_SC_4leg,shengtao1_t1_t2_J_6_8_legs_SC_on_electron_doped_side,shengtao2_8_leg,white_enhanced_SC_electron_side_8leg,dona1_6_8_SC_both_electron_hole_doped,dona2,yifan_6leg_SC_electron_doped_side,yifan2_6leg_SC_electron_side,tJ_4l3g,Hubbard_4leg_noSC}. 
Moreover, a solution to the Hubbard model with intermediate/strong interactions is beyond the reach of most theoretical frameworks based on perturbative approaches. Variational approaches have been used in early research to study the static properties, however they mainly focus on the original Hubbard Model without NNN hopping terms. \cite{ANDERSON1973153,anderson1987resonating, gutzwiller1,gutzwiller2,Baeriswyl_1980,baeriswyl2000variational,balazs} In the regime that both NNN hopping and strong interactions are important, numerical methods 
provide the most reliable means to investigate the low energy physics with quantum fluctuations in a non-perturbative manner.
Our goal in this article is to review the recent progress in numerical studies on the Hubbard model with NNN hopping, also mentioning early work that already unveiled a variety of surprises in this context dating back to decades ago.

The scope of this article will be as follows:
we first introduce the models of interest, then review the recent progress on the effects of the NNN hopping on the Hubbard model, with special focus on the influence on the magnetism and superconductivity. Specifically, we focus on studies using numerical techniques, such as density matrix renormalization group (DMRG), \cite{white1992,white1993,white2004} quantum Monte Carlo (QMC) \cite{DQMC_method1,DQMC_method2,afqmc,qmc_1,qmc_2}, tensor networks (TN) \cite{tn_1,tn_2}, and exact diagonalization (ED)/Lanczos on small systems \cite{dagotto1996surprises,Elbio_2D_t1_t2_Nagaoka_FM}.
The lattice geometries of our focus will be one-dimensional chains and two-dimensional square lattices, including ladder and cylindrical systems as intermediate geometries from one- to two-dimensions. 
We limit our discussions to the quantum regime at zero temperature (or the quantum fluctuations at low temperatures for QMC studies), mainly centering on the ground state properties but commenting on the photoemission spectra as complementary discussions.

\section{Models}

The Hubbard model was initially proposed to describe correlated electrons in periodic lattices \cite{hubbard1963electron}; it is defined as:
\begin{equation}
H_{Hubbard}  =  -t\sum_{\mathbf{r},\sigma}(c_{\mathbf{r},\sigma}^\dagger c_{\mathbf{r}+\hat{\delta},\sigma} + h.c.) +  U\sum_{\mathbf{r}}n_{\mathbf{r},\uparrow} n_{\mathbf{r},\downarrow}
\label{hubbard}
\end{equation}
where $c_{\mathbf{r},\sigma}^\dagger$ ($c_{\mathbf{r},\sigma}$) is the electron creation (annihilation) operator at position $\mathbf{r}$ with spin index $\sigma = \uparrow,\downarrow$, $n_i$ is the electron number operator, $t$ is the hopping integral between nearest neighbors, and $U$ is the on-site Coulomb interaction.
In the strongly interacting limit, the Hubbard model can be mapped to an effective $t-J$ model \cite{mapping_hubbard_tJ}:
\begin{eqnarray}
H_{t-J} & = & -t\sum_{\mathbf{r},\sigma}(c_{\mathbf{r},\sigma}^\dagger c_{\mathbf{r}+\hat{\delta},\sigma} + h.c.)
     +  J\sum_{\mathbf{r}}(\vec{S}_\mathbf{r}\cdot \vec{S}_{\mathbf{r}+\hat{\delta}} 
    - \cfrac{1}{4} n_in_{i+1})
\label{tJ}
\end{eqnarray}
where $\vec{S}_\mathbf{r}$ is the spin $S=1/2$ operator at position $\mathbf{r}$, and $J$ ($=4t^2/U$ when $U$ is large) parametrizes the magnitude of the spin exchange. The strong and weak interactions are differentiated by the value of $J$, and the crossover is around $J=1/4$. This model forbids double-occupancy and captures the physics of the Hubbard model on the hole doped side. Note that the canonical transformation from Hubbard to the $t-J$ model also generates a three-site term $-J/4 \sum_{\langle i,j\rangle, \langle i,j^{\prime}\rangle , j\neq j^{\prime},\sigma} (c_{j^{\prime},\sigma}^\dagger n_{i,-\sigma} c_{j,\sigma} - c_{j^{\prime},\sigma}^\dagger c_{i,-\sigma}^\dagger c_{i,\sigma} c_{j,-\sigma})$, of which the effects are much less studied \cite{KAChao_1977,WY_tJ3s,jia2014persistent,Lema1996,Simon1993,Simon1995,Batista1997,Ammon1995}.

\begin{figure}
\centering
\includegraphics[width=1.0\textwidth]{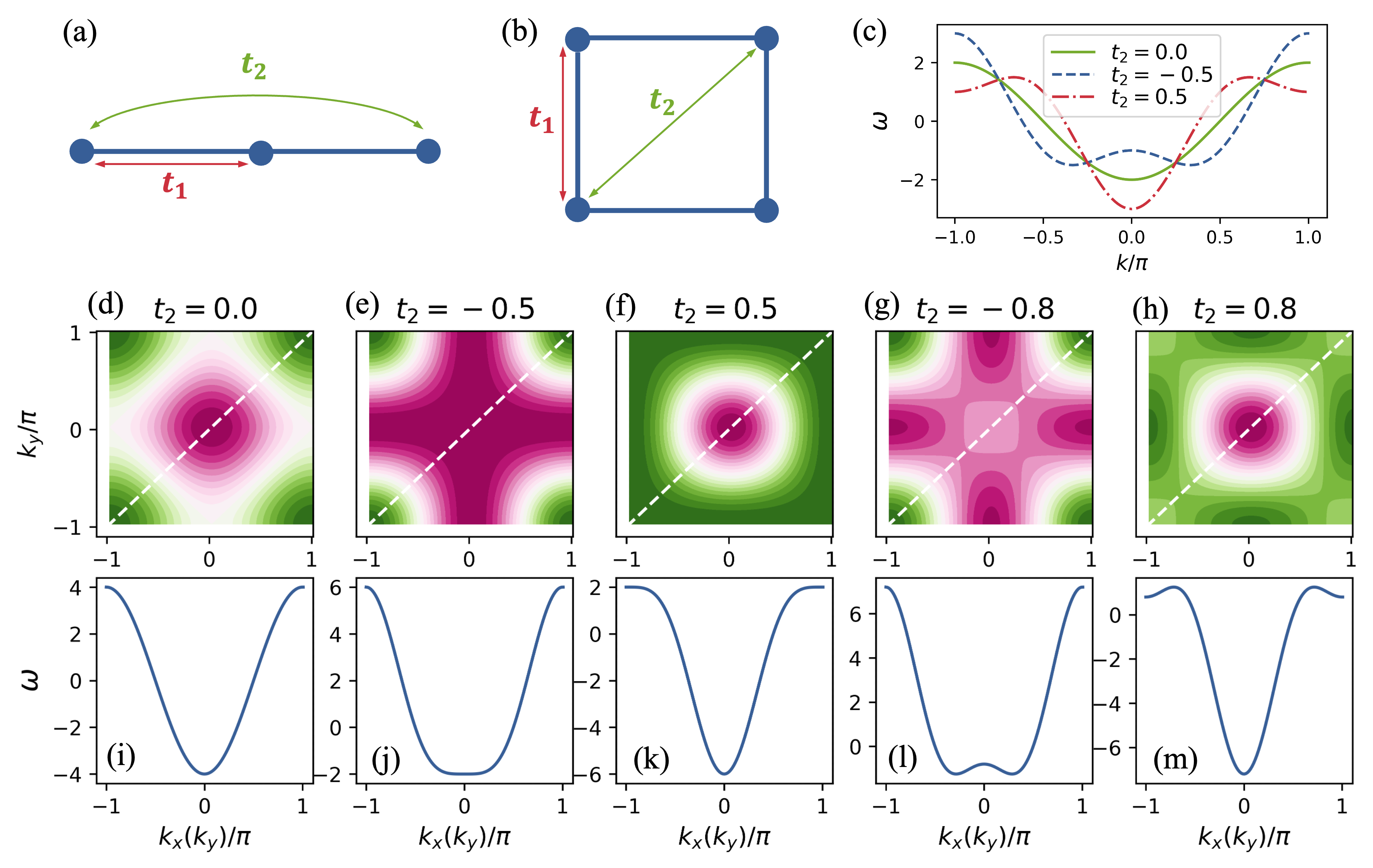}\hfill
\caption{Lattice structures and non-interacting band dispersions. (a) one-dimensional lattice; (b) two-dimensional square lattice; (c) non-interacting bands for one-dimensional chains; (d)-(h): two-dimensional energy dispersion colormaps; (i)-(m): energy dispersion along the dashed-white line cuts in (d)-(h).}
\label{fig:structure}
\end{figure}

By including the NNN hopping, the Hubbard model yields 
the $t_1-t_2-U$ model:
\begin{equation}
    H_{t_1-t_2-U}  =  -t_1\sum_{\mathbf{r},\sigma}(c_{\mathbf{r},\sigma}^\dagger c_{\mathbf{r}+\hat{\delta},\sigma} + h.c.)  -   t_2\sum_{\mathbf{r},\sigma}(c_{\mathbf{r},\sigma}^\dagger c_{\mathbf{r}+\hat{\delta^{\prime}},\sigma} + h.c.)  +  U\sum_{\mathbf{r}}n_{\mathbf{r},\uparrow} n_{\mathbf{r},\downarrow}
\label{t1t2U}
\end{equation}
where $\hat{\delta}$ ($\hat{\delta^{\prime}}$) is the vector pointing to the nearest (next-nearest) neighbors, and $t_1$ ($t_2$) is the hopping parameter between nearest (next-nearest) neighbors (Fig.\ref{fig:structure} (a) and (b)). The definition of ``next-nearest neighbor'' differs between one-dimensional chains and two-dimensional square lattices. In the former case, the hopping of the system transforms the chain into a ``zigzag'' ladder, while in the latter case, the NNN hopping is along the diagonal of each cell, as shown in Fig.\ref{fig:structure}. 
In the strongly interacting limit, the $t_1-t_2-U$ model can be mapped to the ``$t_1-t_2-J_1-J_2$'' model:
\setlength{\mathindent}{0cm}
\begin{eqnarray}
\nonumber H_{t_1-t_2-J_1-J_2}  =  -t_1\sum_{\mathbf{r},\sigma}(c_{\mathbf{r},\sigma}^\dagger c_{\mathbf{r}+\hat{\delta},\sigma} + h.c.)  -  t_2\sum_{\mathbf{r},\sigma}(c_{\mathbf{r},\sigma}^\dagger c_{\mathbf{r}+\hat{\delta^{\prime}},\sigma} + h.c.) \\ \label{t1t2J}
 \ \ \ \ \ \ \ \ \ \ \ \ \ \ \  + J_1\sum_{\mathbf{r}}(\vec{S}_\mathbf{r}\cdot \vec{S}_{\mathbf{r}+\hat{\delta}} 
    - \cfrac{1}{4} n_{\mathbf{r}}n_{\mathbf{r}+\hat{\delta}}) + J_2\sum_{\mathbf{r}}(\vec{S}_\mathbf{r}\cdot \vec{S}_{\mathbf{r}+\hat{\delta^{\prime}}}
    - \cfrac{1}{4} n_{\mathbf{r}}n_{\mathbf{r}+\hat{\delta^{\prime}}})\\ \nonumber
\end{eqnarray}
where $J_1=4t_1^2/U$ and $J_2=4t_2^2/U$ when $U$ is large. A similar $t_1-t_2-J_1-J_2$ model can also be obtained by a direct simplification of the Emery model \cite{ESKES1989424,Eskes1991,Eskes1991_2}. Because the $J_2$ term is roughly an order of magnitude smaller than $J_1$, it is often omitted and the $t_1-t_2-J$ model, where $J$ is the $J_1$ term in Eq.\ref{t1t2J}, is frequently studied.

Hereinafter, we refer to the ``Hubbard model'' when we focus on the original nearest-neighbor (NN) only Hubbard model, and ``$t_1-t_2-U$'' model when we are describing the Hubbard model with NNN hopping.

Several different arguments support the necessity of including a longer-range hopping when probing superconducting phases.
One of them is directly motivated by the 
 band structure of the cuprates, which are usually considered as layered materials, where each layer can be depicted as a Lieb lattice composed of the copper $3d_{x^2-y^2}$ orbital and oxygen $2p_\sigma$ orbitals \cite{PWAnderson_book}. Based on this picture, a three-band Hubbard model -- also referred-to as ``Emery model'', was proposed as the minimal theory to describe the low energy physics of cuprate materials \cite{Emery_1,VARMA1987681}.
Subsequently, it was argued that the Emery model can be simplified to an effective single band Hubbard model. This mapping has been illustrated by (i) the Zhang-Rice (ZR) singlets argument, which demonstrated that the dynamics of the ZR singlets in the Emery model is equivalent to the dynamics of the holes in the single band Hubbard model \cite{ZRS}; (ii) the level splitting calculations, which proved the equivalence of energy levels between Emery model and a single-band model \cite{ESKES1989424,Eskes1991,Eskes1991_2}.
Furthermore, the band structures obtained from first principle calculations also support the suggestion that a single-band Hubbard mode with NNN hopping should suffice in describing some low energy aspects of high-Tc cuprates \cite{DFT_1,DFT_2}.
The hopping parameters can also be acquired from the photoemission spectroscopy results \cite{read_off_t2_photoemission_experiment} although it was argued to be not always reliable \cite{Moreo1995}.

\section{One-dimensional Hubbard model with NNN hopping}

Recent experimental studies unveiled the particle-hole asymmetry of the one-dimensional cuprate material Ba$_{2-x}$Sr$_x$CuO$_{3+\delta}$. \cite{chen2021anomalously} Numerical investigations on this material suggest that the NNN hopping, the three-site hopping and the nearest-neighbor attractions all contribute to the $3k_F$ and holon-folding branches in the photoemission spectrum. \cite{AdrianAligia}. The accurate modeling of these relatively simple cuprate materials, and the understanding of the role of the different terms, can shed light on the mechanisms responsible for hole pairing in their higher-dimensiona counterparts.
The low-energy physics of the Hubbard and $t_1-t_2-U$ models on a chain is by now well established. The one-dimensional Hubbard model can be exactly solved using Bethe Ansatz \cite{bethe,essler2005one}. It is a Mott insulator at half-filling and always exhibits Luttinger Liquid behavior upon doping. In one-dimension, Landau's quasiparticles are never realized, and the spectrum displays edge singularities instead of Lorentzian peaks near the Fermi energy. Furthermore, the Hubbard model manifests a phenomenon known as spin-charge separation, in which the natural excitations split into those carrying spin, and those carrying charge, with independent velocities and characteristic energy scales \cite{kim1996observation,kim2006distinct,auslaender2005spin,jompol2009probing}. In one-dimension, spontaneous symmetry breaking of continuous symmetries is strictly forbidden, and correlations decay either algebraically or exponentially with distance. Therefore, superconductivity is defined as a regime where the pair-pair correlations become dominant, i.e, decay algebraically with a small power, compared to all other instabilities. In 1D, the Hubbard model never displays superconductivity, even after the addition of NNN hoppings. Nevertheless, the closely related $t-J$ and $t_1-t_2-J$ models exhibit a superconducting phase. In the $t-J$ model, superconductivity emerges at intermediate $J$.

\begin{figure}
\centering
\includegraphics[width=0.7\textwidth]{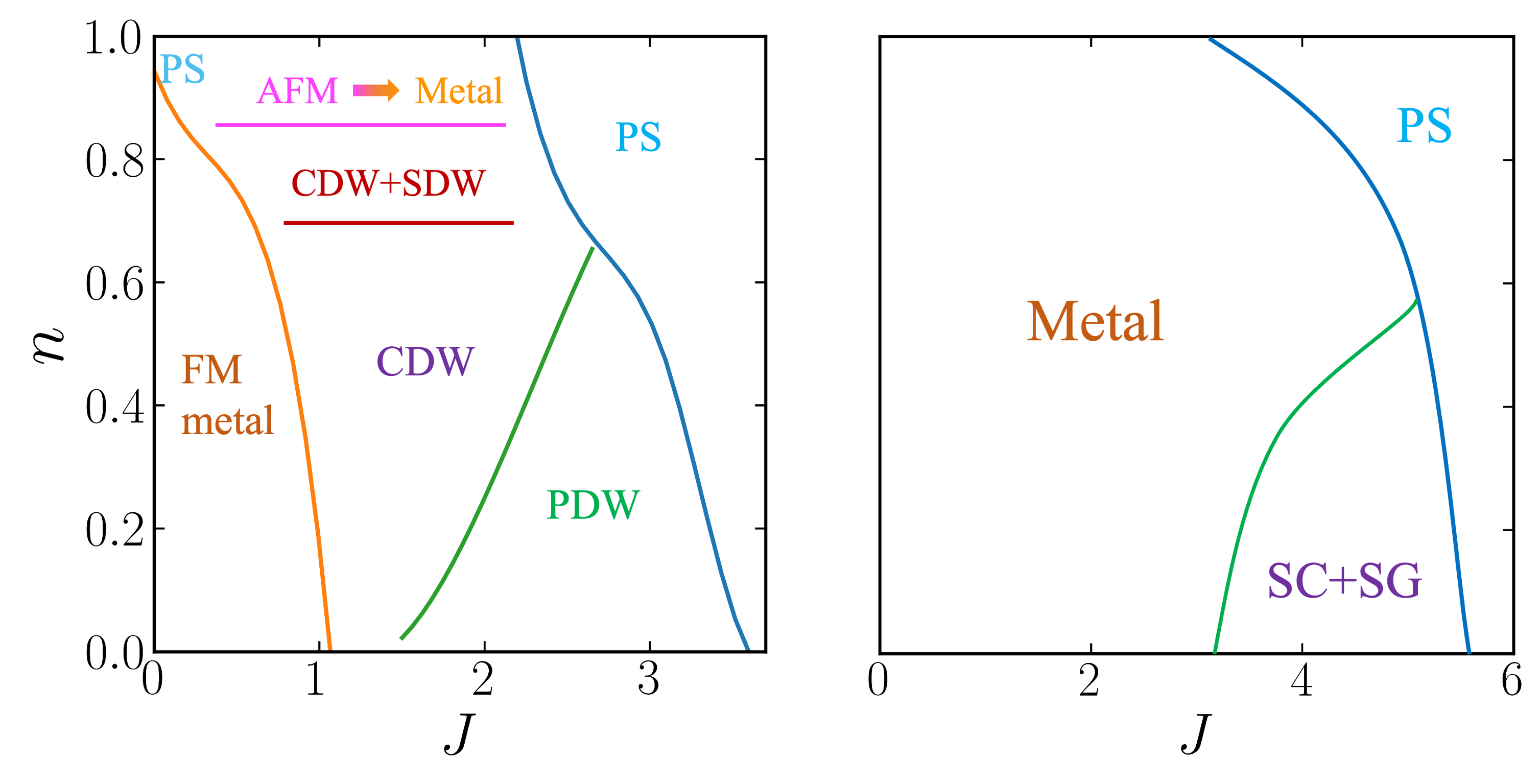}\hfill
\caption{Left: Phase diagram for $t_1-t_2-J$ model with $t_2=-0.5$; Right: Phase diagram for $t_1-t_2-J$ model with $t_2=0.5$. AFM: antiferromagnetic; CDW: charge density wave; SDW: spin density wave; FM: ferromagnetic; PDW: pair density wave; PS: phase separation; SC: superconducting; SG: spin-gapped. \cite{my_t1_t2_J}}
\label{fig:phase1d}
\end{figure}

In this section, we focus on the magnetism and superconductivity in the $t_1-t_2-J$ model in one spatial dimension. This model shows a rich phase diagram as a function of density $n$ and spin exchange $J$ \cite{my_t1_t2_J}. With a negative NNN hopping ($t_2=-0.5$), a spin density wave is realized near half-filling, a fully polarized FM phase occurs at small $J$, which coincides with the FM phase in the $t_1-t_2-U$ model when the Coulomb interaction $U$ is strong \cite{Muller,Hlubina1999,Nishimoto2008}, and  a charge density wave appears in the region with intermediate $J$ upon doping, a pair density wave emerges by further increasing $J$, and the system finally develops phase separation at large $J$ (Fig.\ref{fig:phase1d}) \cite{my_t1_t2_J}. With a positive NNN hopping ($t_2=0.5$), nevertheless, the system resembles the phase diagram of the $t-J$ model qualitatively (Fig.\ref{fig:phase1d}) while the pairing order at intermediate $J$ displays a conventional uniform SC \cite{my_t1_t2_J}. 

In order to illustrate the ground state properties, we show various correlations measured on the ground state wave functions.
The spin-spin correlations are defined as:
\begin{equation}
S(r) = \langle S^z_0S^z_r\rangle;
\label{Sr}
\end{equation}
the density-density correlations as: 
\begin{equation}
D(r) = \langle n_0n_r\rangle - \langle n_0\rangle \langle n_r\rangle;
\label{NNr}
\end{equation}
the singlet pair-pair correlations are defined as:
\begin{equation}
P_s(r) = \langle \Delta^\dagger_0\Delta_r\rangle, 
\label{Psr}
\end{equation}
where $\Delta^\dagger$ operator creates a singlet pair on neighboring sites,
\begin{equation}
\Delta^\dagger_i = \frac{1}{\sqrt{2}}(c^\dagger_{i,\downarrow}c^\dagger_{i+1,\uparrow} - c^\dagger_{i,\uparrow}c^\dagger_{i+1,\downarrow}).
\end{equation}
and the triplet pair-pair correlations are:
\begin{equation}
P_t(r) = \langle \Tilde{\Delta}^\dagger_0\Tilde{\Delta}_r\rangle + \langle c^\dagger_{0,\downarrow}c^\dagger_{1,\downarrow} c_{r,\downarrow}c_{r+1,\downarrow}\rangle + \langle c^\dagger_{0,\uparrow}c^\dagger_{1,\uparrow} c_{r,\uparrow}c_{r+1,\uparrow}\rangle
\label{Ptr}
\end{equation}
where $\Tilde{\Delta}^\dagger$ operator creates a triplet pair on neighboring sites:
\begin{equation}
\Tilde{\Delta}^\dagger_i = \frac{1}{\sqrt{2}}(c^\dagger_{i,\downarrow}c^\dagger_{i+1,\uparrow} + c^\dagger_{i,\uparrow}c^\dagger_{i+1,\downarrow})
\end{equation}
The static structure factors are obtained by Fourier transform the correlations from real space  to momentum space.

\begin{figure}
\centering
\includegraphics[width=1.0\textwidth]{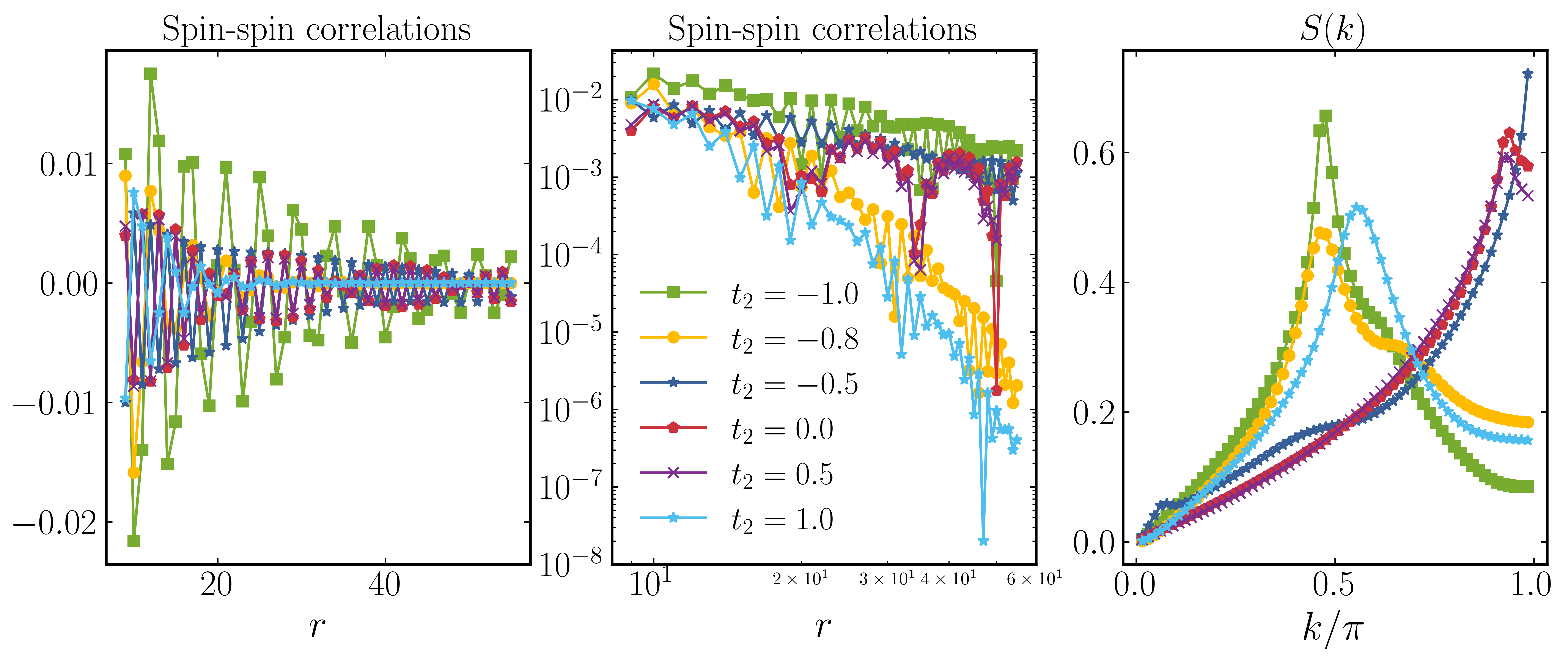}\hfill
\includegraphics[width=.8\textwidth]{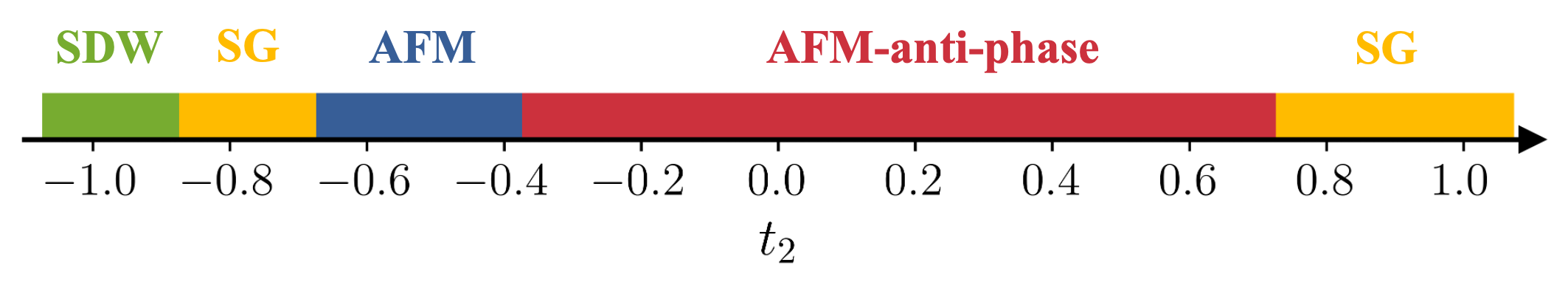}\hfill
\caption{Top: Spin-spin correlations (left and middle) and spin structure factors (right) for the one-dimensional $t_1-t_2-U$ model with various values of $t_2$ and $U=8$. Results are obtained by DMRG for the chains with length $L=64$ and $1/16$ hole doping. Bottom: Magnetic phase diagram as a function of $t_2$ for $1/16$ hole doping, where the green region represents spin density wave, the yellow area depicts spin-gapped phase, the blue zone stands for AFM, and the red sector shows the AFM with anti-phase domain walls. }
\label{fig:szsz_t1t2U}
\end{figure}

\subsection{The influence of NNN hopping on magnetism upon light doping} 

Antiferromagnetism is often considered to be the ``pairing glue'' for superconductivity \cite{pairing_glue1}, and the momentum vector of the magnetic order can determine the oscillation pattern of the pairing orders upon higher doping \cite{kh_pwd,my_pdw}. Therefore, investigating the magnetic properties near half-filling will allow us to understand superconductivity in the optimally doped regime. 

The Hubbard/$t-J$ model at half-filling with strong Coulomb interactions is a Mott insulator, displaying quasi-long-range antiferromagnetic order  \cite{Mermin1966}, and the spin channel behaves similarly as the Heisenberg model \cite{Fradkin_book}. The effect of NNN hopping is minimal at half-filling in the strongly interacting limit, whereas the effect on magnetism becomes noticeable upon light doping, where the system shows particle-hole asymmetry.

This particle-hole asymmetry can be observed by changing the sign of $t_2/t_1$ from positive to negative, which maps the upper Hubbard band into the lower Hubbard band, hence the same model can be used to illustrate low-energy physics on both electron- and hole-doped sides \cite{lee_review}.
The magnetism upon light doping is sensitive to the sign and magnitude of $t_2$ \cite{nnn_1d_photoemission}. In Fig.\ref{fig:szsz_t1t2U} we show the spin-spin correlations and the static spin structure factors for the $t_1-t_2-U$ model for various values of $t_2/t_1$, where we fix $U=8$ and doping density to be $1/16$. 
When $t_2/t_1$ is smaller than -0.9, the spin-spin correlations oscillate as a spin density wave (SDW).
Increasing the value of $t_2/t_1$, there is an ``optimal'' regime ranges from -0.7 to -0.4 where the spin structure factor peaks at $\pi$. An early study on the $t_1-t_2-U$ model \cite{Daul2000} and $t_1-t_2-J$ model \cite{my_t1_t2_J} have also revealed that the AFM order is favored in this regime. The strong tendency to AFM near half-filling for $t_2/t_1 =-0.5$ leads to phase separation \cite{my_t1_t2_J}. 
In a large region when $t_2/t_1$ is between -0.4 and 0.7, the magnetic order is AFM with anti-phase domain walls, which is similar to the stripe phase observed in higher dimensions \cite{antiphase_huang2017numerical}. There are also two regimes when $t_2/t_1$ is between -0.9 and -0.7, and when $t_2>0.7$, where the magnetic order is short-range and the system is spin-gapped (Fig.\ref{fig:szsz_t1t2U}).

\subsection{Superconducting region} 

\begin{figure}
\centering
\includegraphics[width=0.29\textwidth]{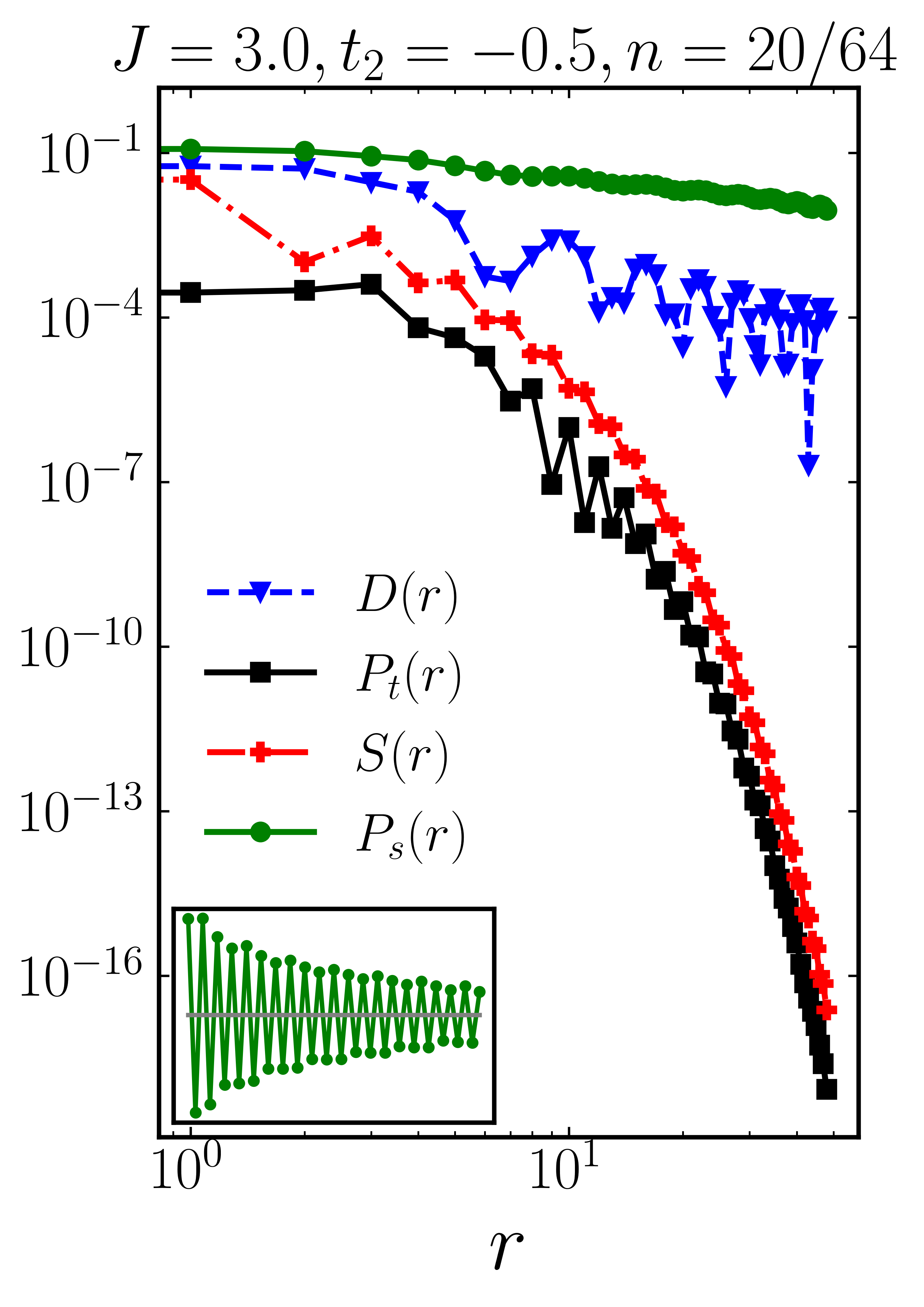}\hfill
\includegraphics[width=0.28\textwidth]{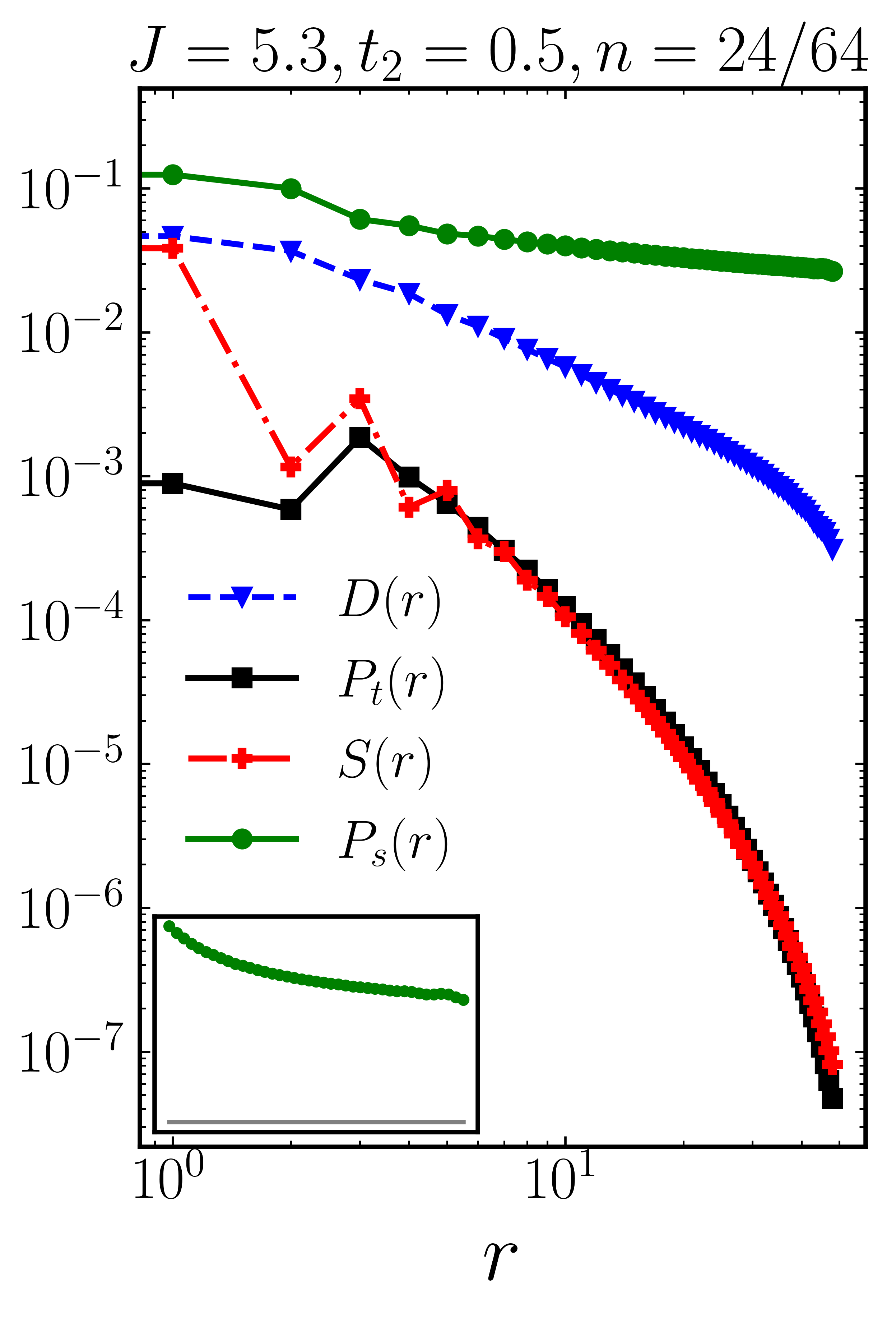}\hfill
\includegraphics[width=0.33\textwidth]{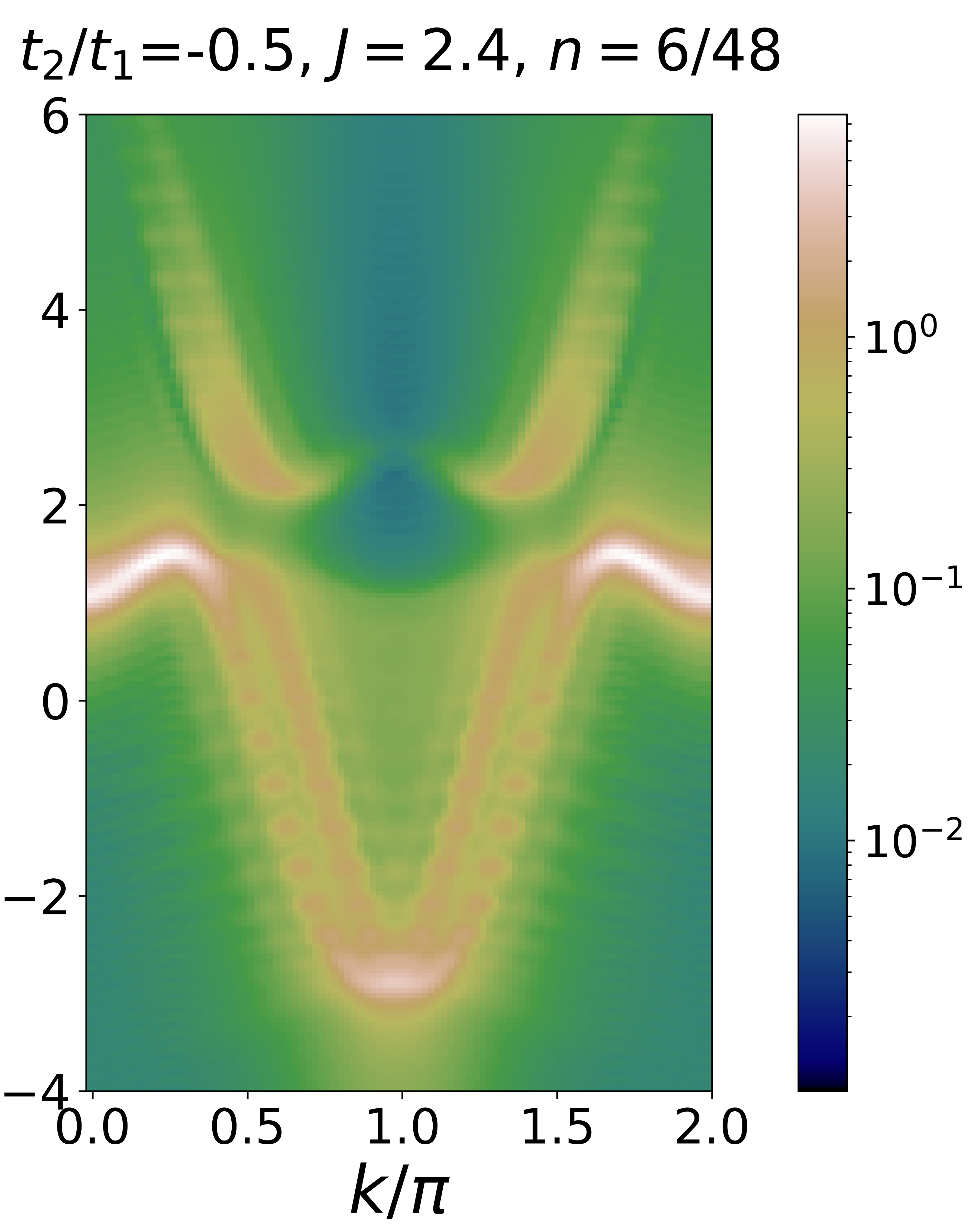}\hfill
\caption{Left: Ground state correlations for the $t_1-t_2-J$ chain with $t_2=-0.5$, $J=3$ and density $n=20/64$, where the system is in the PDW phase. Middle: Correlations for the $t_1-t_2-J$ chain with $t_2=0.5$, $J=5.3$ and density $n=24/64$, where the system is in the SC phase. Insets: Pairing orders shown in a linear scale. \cite{my_t1_t2_J} Right: Photoemission spectrum for the $t_1-t_2-J$ chain with $t_2=-0.5$, $J=2.4$ and $n=6/48$. \cite{my_pdw} }
\label{fig:pairing_tj}
\end{figure}

The original one-dimensional Hubbard model does not accommodate superconductivity.
The inclusion of NNN hopping can frustrate the magnetic order and open a spin gap, which might be favorable for the pairing correlations. However, although some early studies suggested that superconductivity is present in such a system when $t_2$ is in a specific regime \cite{Nishimoto2008,Fabrizio1996}, we found no evidence of a SC phase in this system, according to our DMRG calculations. 
On the other hand, the $t_1-t_2-J$ model exhibits a SC phase in the spin gapped regime with intermediate value of $J$ at small electron densities \cite{my_t1_t2_J}. In Fig.\ref{fig:pairing_tj} we present the pairing correlations in the SC phase for $t_2=0.5$ (left panel) and $t_2=-0.5$ (middle panel). They oscillate as a PDW in the former case, and an uniform SC in the latter (results from Ref.\cite{my_t1_t2_J}). The pattern of the spacial oscillation of the pairing order with $t_2=-0.5$ is also closely associated with magnetism \cite{kh_pwd}, which shows AFM tendency in this regime.

The exotic PDW phase in the $t_1-t_2-J$ model is attributed to the NNN hopping. When $t_2/t_1$ is negative (i.e., -0.5), the topology of the non-interacting band survives under the influence of Coulomb interactions, exhibiting four Fermi points upon doping. 
Evidence of this underlying relation between the onset of PDW and the NNN hopping \cite{my_pdw} can be further found in the momentum resolved spectrum. In the right panel of Fig.\ref{fig:pairing_tj} we present the 
data from Ref.\cite{my_pdw}, which shows the photoemission spectrum for the PDW phase of the $t_1-t_2-J$ model 
\cite{my_t1_t2_J}. Even with the presence of the interaction, the spectrum dispersion still shows two local minima. This second neighbor term allows electrons to hop without frustrating the spin short-range order, thus the AFM tendency is preserved, and the PDW order is favored.

\subsection{Ferromagentism in the one dimensional $t_1-t_2-U$ model}

\begin{figure}
\centering
\includegraphics[width=1.0\textwidth]{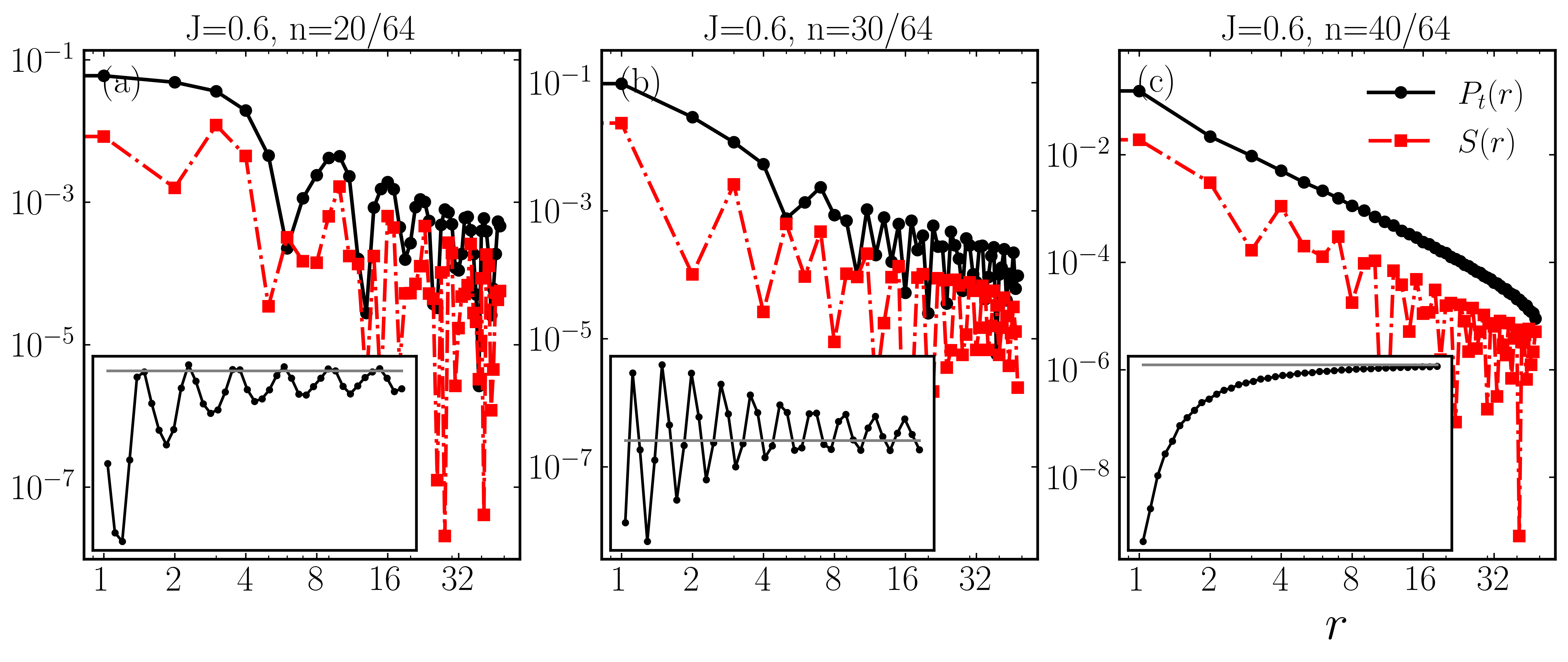}\hfill
\caption{Spin-spin and triplet pairing correlations in the fully polarized phase of $t_1-t_2-J$ model. The correlations are measured for the highest spin sectors, and are shown in a log-log scale. Insets: triplet-SC order shown in a linear scale. \cite{my_t1_t2_J}}
\label{fig:fm_phase_1d}
\end{figure}

The fully polarized FM (FPFM) state of the Hubbard model with a single hole was first proposed by Nagaoka \cite{Nagaoka1966} for systems with infinitely strong on-site Coulomb interactions. 
It was first proved to exist in lattices in two or higher dimensional bipartite lattices
\cite{Nagaoka1966}. Originally, Nagaoka's FM is absent in the one-dimensional Hubbard model. By introducing a negative second-neighbor hopping, Nagaoka’s FM can be extended to the one-dimensional Hubbard model \cite{Mattis1974}. 
The $t_1-t_2-U$ (or $t_1-t_2-J$) model with negative $t_2/t_1$ is one of the few models that has been found to have a FPFM ground state \cite{Muller,Hlubina1999,Nishimoto2008}. 
Efforts have been devoted to the exploration of whether this mechanism can be extended to finite Coulomb interaction and more realistic doping concentrations.
Numerical studies have shown later that, depending on the filling density and the value of $|t_2|$, the fully polarized FM phase could be realized at finite values of $U$ \cite{Daul1998} indicating that both, the ``single hole doping'' and ``infinite $U$'' conditions for Nagaoka FM can be relaxed. 
In fact, Nagaoka FM exists in a large region in the density-exchange phase diagram of the $t_1-t_2-J$ model \cite{my_t1_t2_J}, and strikingly, this fully polarized phase also displays a strong triplet pairing order (Fig.\ref{fig:fm_phase_1d}). Depending on the filling density, this triplet pairing order develops as a PDW, making this model a tunable platform to study triplet-superconductivity.

\section{Two-dimensional Hubbard model with NNN hopping}
 
Angle-resolved photoemission spectroscopy studies have revealed that the two-dimensional cuprate materials such as LSCO and BSCCO exhibit band structures that require NNN hopping. \cite{read_off_t2_photoemission_experiment,cuprate_SC_phase_diragm_experiment}
The two-dimensional Hubbard model and its extended forms are considered to be the minimal models that can describe the physics of the cuprate high-critical-temperature superconductivity. However, accessing the ground state properties of large two-dimensional systems by numerical means remains one of the most pressing challenges in computational condensed matter. \cite{absence_hubbard_2D,raghu_review,qin2022review}. Furthermore these studies are sensitive to system size, lattice geometry, and the boundary conditions of choice. Having a reduced Hilbert space, the $t_1-t_2-J$ model (where $J$ is small) is often studied as a  substitution for the $t_1-t_2-U$ model. Although double-occupancy is forbidden in the $t_1-t_2-J$ model, the effects of electron doping can still be studied through
the particle-hole
transformation by changing the sign of $t_2$. The conclusions for $t_1-t_2-J$ model (when $J$ is small) and the $t_1-t_2-U$ model (when $U$ is large) on magnetic order near half filling, \cite{positive_t2_enhance_AFM_ED,constrained_path, electron_afm_1,antiphase_huang2017numerical,afm_to_fm_2d} the superconductivity \cite{Elbio_Dagotto_1999_tJ_heisenberg_2leg_ladder, my_3band} and the Nagaoka FM \cite{ Elbio_2D_t1_t2_Nagaoka_FM,2D_Nagaoka_FM_nagative_t2, 2D_t_J_breakdown_Nagaoka} are qualitatively the same.

Recent large-scale computations demonstrated that the SC phase is weak in the original Hubbard model \cite{absence_hubbard_2D}. Hence, the clues for superconductivity may be found in other missing terms in the Hamiltonian.
By including the NNN hopping, the $t_1-t_2-U$ model exhibits a rich phase diagram.
In the lightly doped regime, numerical studies have confirmed the AFM pattern near half-filling \cite{antiphase_huang2017numerical,Tmatrix_hole_no_SC, raghu_review,positive_t2_enhance_AFM_ED,electron_afm_1} and superconductivity on the electron-doped side, regardless of the method and the lattice size \cite{Elbio_tJ_2leg_ladder_pairing,Hubbard_2leg,jiang2019sc_4leg_negativet2,Hubbard_positive_t1_t2_4leg,plaqutte_SC_4leg,shengtao1_t1_t2_J_6_8_legs_SC_on_electron_doped_side,shengtao2_8_leg,white_enhanced_SC_electron_side_8leg,dona1_6_8_SC_both_electron_hole_doped,dona2,yifan_6leg_SC_electron_doped_side,yifan2_6leg_SC_electron_side}. 

Nonetheless, discrepancies between the numerical results for the microscopic models and the experimental phase diagrams of cuprates are still non-negligible. Although there is numerical evidence supporting the appearance of a SC phase \cite{hubbard_2d_SC_in_both} on the hole-doped side, doubts remain because there is considerable research suggesting that the $t_1-t_2-U$ model lacks the ``superconducting dome'' on the hole-doped side \cite{cuprate_SC_phase_diragm_experiment}.

In this section, we review the recent numerical progress understanding the effects of NNN hopping in two-dimensional systems. Our discussion of the two-dimensional $t_1-t_2-U$ and $t_1-t_2-J$ models mainly focuses on two regimes: the lightly doped regime where superconductivity arises, and the highly doped regime where flat-band FM emerges.

\subsection{The influence of next-nearest-neighbor hopping on magnetism near half-filling} 

\begin{figure}
\centering
\includegraphics[width=0.8\textwidth]{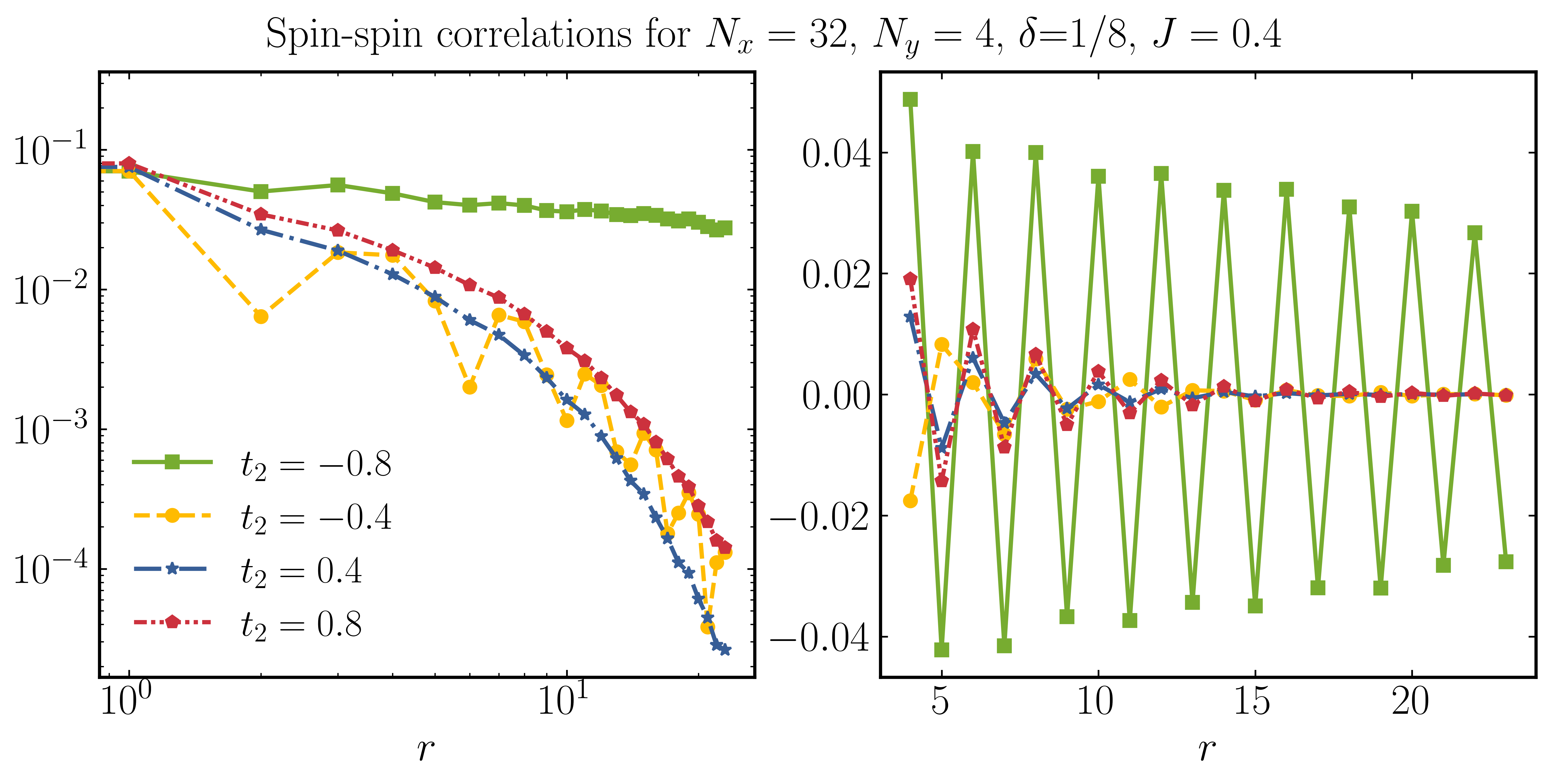}\hfill
\caption{Spin-spin correlations for the $t_1-t_2-J$ model with 1/8 hole doping at fixed $J=0.4$ and various $t_2/t_1$ in a log-log scale (left) and in a linear scale (right). Calculated by DMRG for a cylindrical system with length $N_x=32$ and $N_y=4$.}
\label{fig:afm_tj_4leg}
\end{figure}

The Hubbard model in a two-dimensional square lattice at half-filling is a Mott insulator with long-range AFM (N\'eel) order in the ground state \cite{Mermin1966}.
The AFM order at half-filling is resistant to small values of NNN hopping $t_2$, in fact, when $|t_2|$ is relatively small, i.e., $|t_2| \leq t_1/2$, and $U$ is greater than a threshold, the long-range AFM persists \cite{raghu_review}.
Upon doping, the system lacks particle-hole symmetry, and both the sign and magnitude of $t_2$ affect the magnetism. Numerical studies have found that on the electron doped side (when $t_2 > 0$) the spins form an AFM pattern.
However, on the hole doped side, the spin momentum vector is shifted away from $(\pi,\pi)$ \cite{positive_t2_enhance_AFM_ED,constrained_path, electron_afm_1,antiphase_huang2017numerical,afm_to_fm_2d}.
It is worth noting that this conclusion holds for relative small $|t_2|$ ($<0.5$), which is in the more realistic range that is appropriate for describing the cuprate superconductors. 
By increasing the magnitude of $|t_2|$, both the hole- and electron-doped sides display AFM tendencies upon small doping. Nevertheless, the AFM order on the hole doped side is more robust and exhibits much longer range (Fig.\ref{fig:afm_tj_4leg}).

\subsection{Superconductivity on the electron and hole doped sides}

The superconducting phase of the $t_1-t_2-U$ model has been investigated by various numerical approaches such as DMRG \cite{hubbard_2d_SC_in_both,Elbio_tJ_2leg_ladder_pairing}, QMC \cite{hubbard_2d_SC_in_both,constrained_path}, tensor networks \cite{tensor_nextwork1}, variational Monte Carlo \cite{vMC_negativeT2_SC}, exact diagonalization/Lanczos \cite{Elbio_tJ_2leg_ladder_pairing}, density matrix embedding theory (DMET) \cite{DMET_4_4}, among others\cite{Tmatrix_hole_no_SC,Hlubina_2D_t2_FM_TMA_QMC}. 
Although remarkable progresses have been made by means of these techniques, numerical methods are usually limited to small system sizes or rely on approximations. Therefore, there are still ongoing debates on key issues, such as whether the SC phase in the $t_1-t_2-U$ model coincides with the one observed in experiments. 
As a feasible intermediate step, ladder geometries have been extensively investigated with the aim to extrapolate the two-dimensional physics. Despite the fact that the ladders differ from true two-dimensional system \cite{dagotto1996surprises}, it is immensely valuable to have a quantum system that displays pairing tendencies under control in order to guide our theoretical understanding of the quantum fluctuations in cuprate materials \cite{dagotto1996surprises, rice1996tj,vuletic2006spin}.

\begin{table}
\caption{\label{tabone}Literature survey indicating whether or not ladder systems exhibit a SC phase for the $t_1-t_2-J$ and $t_1-t_2-U$ models.} 
\begin{indented}
\lineup
\item[]\begin{tabular}{@{}*{8}{l}}
\br                              
&&&&&\centre{2}{SC}\cr
\ns
Model&Number&&&&\crule{2}\cr
\  & of legs & $t_2/t_1$ & $J/t_1$ or $U/t_1$ & $\delta$ &  $t_2>0$ & $t_2<0$ & Ref.\cr 
\mr
$t_1-t_2-J_1-J_2$ & \0\0 2 & $[0,0.22]$   & \0 3 & 0.3 & \m\m\m Yes &\m\m\m --- & \cite{Lu_2leg_tJ_ladder}\cr
$t_1-t_2-J$       & \0\0 2 & $[-0.5,0.5]$ &  0.5 & 0.1 & \m\m\m Yes & \m\m\m No & \cite{Elbio_tJ_2leg_ladder_pairing}\cr 
$t_1-t_2-U$       & \0\0 4 & $-0.25$ & \0\0\0\0\0\0 8,12 & $1/8$ & \m\m\m --- & \m\m\m Yes & \cite{jiang2019sc_4leg_negativet2}\cr
$t_1-t_2-U$       & \0\0 4 & $[-0.25,0.25]$ & \0\0\0\0\0\0 8-12 & $[1/12,1/8]$ & \m\m\m Yes & \m\m\m Yes & \cite{Hubbard_positive_t1_t2_4leg}\cr
$t_1-t_2-U$       & \0\0 4 & $-0.25$ & \0\0\0\0\0\0\0 12 & $[1/12,1/8]$ & \m\m\m --- & \m\m\m Yes & \cite{plaqutte_SC_4leg}\cr
$t_1-t_2-J$       & \0\0 6,8 & $[-0.3,0.3]$ &  0.4 & $0-0.25$ & \m\m\m Yes & \m\m\m No & \cite{shengtao1_t1_t2_J_6_8_legs_SC_on_electron_doped_side}\cr
$t_1-t_2-t_3-J$       & \0\0 8 & $[-0.4,0.4]$ &  0.4 & $0.1$ & \m\m\m Yes & \m\m\m No & \cite{shengtao2_8_leg}\cr
$t_1-t_2-J$       & \0\0 5-8 & $\pm 0.2$ &  0.5 & $1/8$ & \m\m\m Yes & \m\m\m No & \cite{white_enhanced_SC_electron_side_8leg}\cr
$t_1-t_2-J_1-J_2$ & \0\0 6,8 & $[-0.22,0]$   &  $1/3$ & $[1/36,1/8]$ & \m\m\m Yes &\m\m\m Yes & \cite{dona1_6_8_SC_both_electron_hole_doped}\cr
$t_1-t_2-J_1-J_2$ & \0\0 8 & $[-0.2,0.3]$   & $1/3$ & $[0.1,0.2]$ & \m\m\m Yes &\m\m\m Yes & \cite{dona2}\cr
$t_1-t_2-U-V$ & \0\0 6 & $\pm 0.4$   & \0\0\0\0\0\0\0 $12$ & $1/12$ & \m\m\m Yes &\m\m\m --- & \cite{yifan_6leg_SC_electron_doped_side}\cr
$t_1-t_2-U$ & \0\0 6 & $[-0.4,0.5]$   & \0\0\0\0\0\0\0 $12$ & $[1/18,1/8]$ & \m\m\m Yes &\m\m\m No & \cite{yifan2_6leg_SC_electron_side}\cr
\br
\end{tabular}
\end{indented}
\end{table}

Two-leg ladders are spin gapped at half-filling, which is favorable for superconductivity. The Hubbard model with strong Hubbard interactions exhibits a Luther-Emery (LE) phase up to $1/8$ hole doping \cite{Hubbard_2leg} and the closely related $t-J$ model on a two-leg ladder also displays superconductivity \cite{Elbio_Dagotto_1999_tJ_heisenberg_2leg_ladder}. By adding NNN hopping, the $t_1-t_2-U$ model manifests strong particle-hole asymmetry. Early studies have shown that the electron-doped side favors SC while the hole-doped sides suppresses pairing order  \cite{Elbio_tJ_2leg_ladder_pairing,Lu_2leg_tJ_ladder}. Alongside this evidence, a recent study revealed that when $J_1$ is large in the $t_1-t_2-J_1-J_2$ model (with NNN exchange), superconductivity is also present on the hole-doped side \cite{my_3band}. These results may partially explain the hole pairing in the ladder compounds such as Sr$_{14-x}$Ca$_x$Cu$_{24}$O$_{41}$ and Sr$_{14}$Cu$_{24}$O$_{41}$. \cite{ladder_experiment,ladder_experiment1,scheie2025cooperpairlocalizationmagneticdynamics}

Different studies have suggested conflicting conclusions or wider ladder systems, even for the original Hubbard model, due to the exponentially large computational complexity. 
A study on four-leg $t-J$ ladders indicates that the SC phase emerges upon doping \cite{tJ_4l3g}. However, the study on the Hubbard model with weaker $U$ (=4 to 8) seem to indicate that the SC phase is absent \cite{Hubbard_4leg_noSC}. By introducing NNN hopping, SC phase arises on both the electron- and hole-doped side \cite{jiang2019sc_4leg_negativet2,Hubbard_positive_t1_t2_4leg}, where the SC on the hole-doped side displays a plaquette-symmetry that is pathological for the four-leg ladder geometry \cite{plaqutte_SC_4leg}, while the SC on the electron-doped side has $d$-wave symmetry.

Further increasing the width of the ladders, more conflicting results appear. Although strong evidence indicating SC on the electron-doped side of the $t_1-t_2-U$ model is uncovered, it is still unclear whether it also occurs on the hole-doped side, as in the cuprate superconductor phase diagrams \cite{shengtao1_t1_t2_J_6_8_legs_SC_on_electron_doped_side,shengtao2_8_leg,white_enhanced_SC_electron_side_8leg,dona1_6_8_SC_both_electron_hole_doped,dona2,yifan_6leg_SC_electron_doped_side,yifan2_6leg_SC_electron_side}. 
Table~\ref{tabone} offers a literature survey on some important works studying the existence of superconductivity in these models. As one can observe, the results are strongly sensitive to system size, aspect ratio (number of legs), boundary conditions, and even technical details such as the symmetry group that has been used \cite{dona1_6_8_SC_both_electron_hole_doped,dona2}. Therefore, extrapolating any conclusions to true two-dimensions should be done with extra caution. 

\text

\subsection{Ferromagentism in the two-dimensional $t_1-t_2-U$ model}

As in the one-dimensional case, the introduction of NNN hopping also opens the possibility for a spin polarized phase.
There are two types of polarized states in the two-dimensional $t_1-t_2-U$ model on the square lattice: (i) the Nagaoka FM in the strongly interacting limit upon doping with a single hole, and (ii) the flat-band FM near the ``van Hove filling'' \cite{Hirsch_2D_FM,Katanin_2D_hubbard_t2negative_FM,Katanin_2,Hlubina_2D_t2_FM_TMA_QMC,Hlubina_2_PQMC,Honerkamp_FM_2D_RG,Igoshev_2D_FM_RG_DMFT,afm_to_fm_2d}. While Nagaoka FM emerges in the large $U$ limit \cite{Nagaoka1966}, flat-band FM occurs even with at weak Coulomb interaction $U$ as long as the doping is optimal \cite{tasaki1998nagaoka}.

The fully polarized FM phase can be identified by measuring the total spin $\langle S_z \rangle$, which involves less numerical uncertainty than the two-point correlation measurement. In addition, electrons in the fully polarized FM states are equivalent to spinless fermions, reducing the dimension of the effective Hilbert space, thus reducing the computational expense. Moreover, numerical results on the FM phase suffer much less from finite-size effect compared to other cases, and the low-energy physics of FM phases can be probed even on small clusters \cite{my_triangle}. 

As a consequence, our understanding of the origin and mechanism for ferromagnetism in two-dimensions is well established. 
In the two-dimensional $t_1-t_2-J$ ($t_1-t_2-U$) model, similar to the one-dimensional cases \cite{my_t1_t2_J}, a negative NNN hopping stabilizes the Nagaoka FM to both larger $J$ (smaller $U$) and higher doping concentrations \cite{Elbio_2D_t1_t2_Nagaoka_FM,2D_Nagaoka_FM_nagative_t2, 2D_t_J_breakdown_Nagaoka}.
Near the ``van Hove filling'', the negative NNN hopping creates flatness near the bottom of the non-interacting band, and a moderate on-site Coulomb interaction will split out the polarized state to a lower energy.

The mechanism for triplet, or $p$-wave, superconductivity in two-dimensions is less clear.
Although numerical studies have found $p$-wave pairing in the proximity of the flat-band FM phase, e.g., when $t_2/t_1=-0.7$ and with high doping concentrations \cite{Yang_2021_DQMC_triplet}, one could argue that the ``pairing glue'', as the AFM for the singlet-SC, is lacking in the polarized state. Whether or not the triplet-SC dominates over other orders in the vicinity of the FM phase, or is intertwined with other instabilities, as the case in the one-dimensional $t_1-t_2-J$ model, is yet to be confirmed.  \newline

For completeness, note that $t-J$ models with NNN and next-to-next-nearest-neighbor (NNNN) hoppings have also been explored with regard
to the presence or absence of ``stripe'' formation, a topic of much discussion in cuprates. 
For work on this topic, one can consult \cite{PRB59R11649, DEVEREAUX20251354683} and references therein. Even more exotic, numerical evidence
has been gathered suggesting that the quasiparticle weight of one hole can drop to nearly zero by varying the NNN strength in two-dimensional lattices.
A cartoon displaying this physics can be found in Fig. 2b of Ref.~\cite{PRB60R3716} where the concept of ``across the hole'' singlets is discussed, strongly suggesting ``spin-charge separation'', an aspect barely touched
in two-dimensional Hubbard and $t-J$ models that merits more research (see also Refs.~\cite{PRB63014414,PRL845844}).

Also, it is interesting to observe that materials other than the cuprates, such as the iron-based superconductors,
need NNN hoppings in the microscopic descriptions due to their atomic configurations 
\cite{PRB82104508,PRB85024532}.

\bibliographystyle{unsrt}
\section{Conclusions}
The numerical study of the interplay between NNN hopping and electron interactions in Hubbard-type models has offered unique insights for understanding the relation between magnetism and superconductivity. Despite ongoing debates on whether the SC phase can be stabilized on the hole-doped side of the Hubbard model, 
numerical methods have witnessed pronounced progress, owing to the rapid advancement in computational power and algorithms during the last few decades. Much progress has also been made in exploring intertwined orders in the doped regime, determining the optimal conditions for stabilizing SC and exotic pairing phases, and realizing fully polarized FM states. \cite{raghu_review,qin2022review}

For the one-dimensional $t_1-t_2-U$ model, a negative NNN hopping induces a fully polarized FM phase in the strong interaction limit, and this FM order is accompanied by triplet superconductivity \cite{my_t1_t2_J}. 
For the $t_1-t_2-J$ model, this NNN hopping drives the evolution of the uniform SC order into a more exotic PDW order closely associated with the Fermi surface geometry and the underlying magnetic order \cite{my_pdw,my_t1_t2_J}. 
In two-dimensional square lattices, the introduction of the NNN hopping enhances pairing instabilities, introducing superconductivity in the phase diagram. 
On the electron-doped side, various numerical and theoretical approaches have proved the existence of the SC phase. Nevertheless, on the hole-doped side, the debate on the existence of a SC phase is still ongoing. 

NNN hopping also makes the FM phase more favorable in the sense that: (i) it stabilizes the Nagaoka FM in the original Hubbard model (or $t-J$ model) at finite values of $U$ (or larger $J$); and (ii) it induces a new fully polarized FM phase in the proximity of a van Hove singularity.

The study of the effects of NNN hopping on single-band Hubbard models is informative for the study of longer-range hopping in multi-bands Hubbard models,
because the single band Hubbard model can describe the effective low energy physics of multi-bands Hubbard models in some parameter regimes \cite{Emery_1,VARMA1987681,ESKES1989424,Eskes1991,Eskes1991_2,Belinicher1994b,Hirayama2019,AdrianAligia,twoband,Schuttler1992,Zhang1989,Belinicher1994,Aligia1994,Hamad2018,Jiang2020,Aligia2020,Sheshadri2023}. In fact, the longer-range hopping has been found to favor the SC phase in the three-band Hubbard model \cite{my_3band}.
Our understanding of the Hubbard model on square lattices -- the relevance between the Fermi surface topology and the onset of PDW, the SC phase induced by NNN hopping, and the FM phase originated from narrow bands and strong interactions -- not only allows us to peek into the mysteries of high-temperature superconductivity, but can also guide the exploration of intertwined orders and exotic quantum phase transitions in frustrated materials \cite{Jin2022}.

\section{Acknowledgment}

We thank Hong-Chen Jiang and Steven A. Kivelson for fruitful discussions. LY and ED are supported by the U.S. Department of Energy (DOE), Office of Science, Basic Energy Sciences (BES), Materials Sciences and Engineering Division. AEF is supported by the National Science Foundation under grant
No. DMR-1807814.  TPD acknowledges support from the Department of Energy, Office of Science, Basic Energy Sciences, Materials Sciences and Engineering Division, under Contract DE-AC02-76SF00515.

\bibliography{ref}
\end{document}